\documentclass[%
 reprint,
 amsmath,amssymb,
 aps,
floatfix
]{revtex4-2}

\pdfoutput=1
\usepackage{graphicx}
\usepackage{dcolumn}
\usepackage{bm}
\usepackage[utf8]{inputenc}
\usepackage[english]{babel}
\usepackage{xcolor}
\usepackage{natbib}
\usepackage{amsmath}
\begin{document}

\preprint{APS/123-QED}

\title{Energetics and mixing efficiency of lock-exchange gravity currents using simultaneous velocity and density fields}

\author{Partho Mukherjee}
 \affiliation{Geophysical \& Multiphase Flows Laboratory, Department of Mechanical Engineering, Indian Institute of Technology Bombay, India}
 
\author{Sridhar Balasubramanian}%
 \email{sridharb@iitb.ac.in}
\affiliation{Geophysical \& Multiphase Flows Laboratory, Department of Mechanical Engineering and IDP in Climate Sciences, Indian Institute of Technology Bombay, India
}%

\date{\today}

\begin{abstract}
A series of laboratory experiments on energy conserving gravity currents in a lock-exchange facility are conducted for a range of Reynolds numbers, $Re= \frac{U_Fh}{\nu} =$ 485-12270, where $U_F$ is the front velocity of the current, $h$ the current depth, and $\nu$ the kinematic viscosity of the fluid. The velocity and density fields are captured simultaneously using a PIV-PLIF system. A moving average method is employed to compute the mean field and a host of turbulence statistics, namely, turbulent kinetic energy ($K$), shear production ($P$), buoyancy flux ($B$), and energy dissipation ($\epsilon$) during the slumping phase of the current. The subsequent findings are used to ascertain the quantitative values of mixing efficiency, $Ri_{f}$, Ozmidov length-scale ($L_O$), Kolmogorov length-scale ($L_\kappa$), and eddy diffusivities of momentum ($\kappa_m$) and scalar ($\kappa_\rho$). Two different forms of $Ri_{f}$ are characterized in this study, denoted by $Ri_{f}^I=\frac{B}{P}$ and $Ri_{f}^{II}=\frac{B}{B+\epsilon}$. The results cover the entire diffusive regime (3 $<Re_b<$ 10) and a portion of the intermediate regime (10 $<Re_b<$ 50), where $Re_b=\frac{\epsilon}{\nu N^2}$ is the buoyancy Reynolds number that measures the level of turbulence in a shear-stratified flow with $N$ being the Brunt-V$\ddot{a}$is$\ddot{a}$l$\ddot{a}$ frequency. The depth averaged turbulence quantities, $\overline{P}(z)$, $\overline{B}(z)$, and $\overline{\epsilon}(z)$, show a marked increase at the interface of the ambient and current fluids, owing to the development of a shear-driven mixed layer. Based on the changes in the turbulence statistics and the length scales, it is inferred that the turbulence decays along the length of the current. The mixing efficiency monotonically increases in the diffusive regime ($Re_{b}<$10), and is found to have an average value of $\overline{Ri_{f}^{I}}\approx$ 0.15 and $\overline{Ri_{f}^{II}}\approx$ 0.2 in the intermediate regime. Using the values of $Ri_{f}$, the normalized eddy diffusivity of momentum is parameterized as $\frac{\kappa_m}{\nu.Ri_{g}}$=1.2$Re_{b}$, where $Ri_{g}$ is the gradient Richardson number, and normalized eddy diffusivity of scalar as $\frac{\kappa_{\rho}}{\nu}$=0.2$Re_{b}$.

\end{abstract}

\maketitle


\section{\label{sec:level1}Introduction}
Turbulent mixing in a shear-stratified environment is a complex process, the understanding of which is important for natural and industrial flows. Distribution of nutrients and suspended matter in the environment, ocean overflows, katabatic flows, river discharges, and pyroclastic flows are a few examples. Mixing in a shear-stratified flow occurs through exchange of fluids, triggered by Kelvin-Helmholtz (K-H) instability, which leads to a final stable state, but not before going through a series of mixing transitions ({\citet{dimotakis_2000}}, {\citet{doi:10.1063/1.5023033}}). The stratification either suppresses or enhances turbulence and mixing, depending on the configuration of the system. The interplay of shear and density stratification modifies the dynamics of a turbulent flow, which has been a topic of research for a long time now.

A gravity current, defined as the flow or intrusion of a dense fluid into lighter fluid, is one such example of a shear-stratified flow that is driven by pressure or density gradients in the direction of the flow. For a gravity current,  the Boussinesq approximation is often valid, except when the density difference between the fluids is large. The resultant mixing of the two fluids and the eddy diffusivities depends on the level of turbulence and are independent of the fluid properties. The first theoretical attempt to study the propagation of a gravity current was made by Von Karman, who approximated the front velocity as $U_{F}\propto\sqrt{2g'h}$, where $g'=g\frac{\rho_{2}-\rho_{1}}{\rho_{0}}$ is the reduced gravity, with $\rho_{2}$ being the density of the heavier fluid, $\rho_{1}$ the lighter fluid, $\rho_{0}$ the reference density taken to be the mean of the two densities, and $h$ is the depth of the gravity current ({\citet{huppert_2006}}). Later, \citet{benjamin_1968} postulated the dynamics of lock-exchange gravity current using inviscid fluid theory and estimated $h$ and $U_{F}$. Different regions of a gravity current based on $h$, $U_{F}$, and associated dynamics were first proposed by {\citet{britter_simpson_1978}} and later revised by {\citet{simpson_britter_1979}}; {\citet{britter_linden_1980}}; {\citet{huppert_simpson_1980}}. For a lock-exchange current, two distinct regions have been documented: (i) a head having a foremost point (front/nose) slightly lifted from the surface, and is the transient portion where the front moves with a constant velocity $U_{F}$ and (ii) a body composing the major portion of the gravity current. These regions are usually distinguished based on the fractional depth, $h/H$ (\citet{doi:10.1146/annurev.fl.14.010182.001241}), where $H$ is the depth of the ambient fluid. The head's frontal view will usually exhibit lobe and cleft structures that form because of the no-slip condition imposed by the surface lying underneath the current (\citet{simpson1969}) causing a certain amount of mixing. As the head progresses, it displaces the lighter ambient and engulfs or ``entrains" it within itself, finally feeding the mixed fluid to the stratified body (\citet{sher_woods_2015}). The body of the current, on the other hand, undergoes mixing through Holmboe waves or Kelvin-Helmholtz instability. The gravity current once released from its initial lock-exchange position undergoes a series of transition (\citet{cantero_lee_balachandar_garcia_2007}). First there is a slumping phase, where the heavier fluid  ``slumps" and forms a gravity current. The front maintains a constant velocity during the slumping phase and the current is energy conserving. As the heavier fluid slumps, the lighter fluid rushes behind to fill the void, and in this process a wave from the end wall is reflected, which is called the ``bore''. The bore catches up with the front and once that happens, it marks the end of the slumping phase. The spatial location at which the bore catches up with the front is known as the slumping point. The slumping point usually lies at around 5-10 lock lengths from the gate, depending on the initial configuration (\citet{rottman_simpson_1983}). Immediately after this, the current undergoes an inertial phase, where the buoyancy and inertial forces are in balance. In the inertial phase, though the front velocity now decays in a self similar manner (significant dilution of the current and it is no more energy conserving ({\citet{sher_woods_2015}})), the body of the gravity current has a quasi-steady behavior, which has been quantified through the measurements of mean velocity and density fields (\citet{Chowdhury2014}). This is followed by a purely viscous phase (where buoyancy and viscous forces are in balance) where the current retards significantly and the turbulence almost vanishes. If the viscosity plays a role in the slumping phase itself, the current skips the inertial phase (\citet{huppert_simpson_1980}).

The earliest account on bulk mixing in a gravity current was reported by \citet{ellison_turner_1959}. Referred to as an inclined plume in their research, that had a negatively buoyant source as the driving force, the entrainment coefficient was based on the velocity at which the ambient entrained into the mean flow. Thus, entrainment as a function of an overall Richardson number, $Ri = g(\rho - \rho_a) h/\rho_a V^2$ was proposed, where $g$ is the acceleration due to gravity, $h$ is the moving layer's depth, $\rho_a$ and $\rho$ are the ambient density and the moving layer's density, finally $V$ is the velocity of the moving layer. Subsequently, a link between the type of instability and mixing was given by  {\citet{doi:10.1080/03091927908243758}}. The qualitative observations pertaining to mixing in lock-exchange kind of flows were also made by \citet{HACKER1996183} and \citet{hallworth_huppert_phillips_sparks_1996}. A $Ri$ based power law was proposed for parameterizing entrainment in a sloping gravity current ({\citet{princevac_fernando_whiteman_2005}}), where the argument was that the entrainment law proposed by \citet{ellison_turner_1959} underestimated the entrainment rate. The entrainment and mixing in a lock-exchange gravity current, based on the available potential energy method, was reported for the slumping phase by \citet{fragoso_patterson_wettlaufer_2013} using light attenuation method. The experimental density distribution of fluid elements were cross-sorted using the framework used by  \citet{winters_lombard_riley_d'asaro_1995} to redistribute the density field in its minimum potential energy state and the entrainment and mixing was formulated based on the evolution of available potential energy and the background potential energy. More recently, mixing efficiency (a detailed explanation of which is presented in \S\ref{TKE}) in a lock-exchange setup has been reported for a range of Reynolds numbers, $Re=\frac{U_{F}h}{\nu}$, where $U_{F}$ is the front velocity of the current and $h$ is the current depth. Firstly, \citet{ILICAK20141} reported the mixing efficiency for $125<Re<10000$ using DNS based on the the evolution of background potential energy. Experimentally, \citet{hughes_linden_2016} for $7000<Re<72000$, and \citet{micard:hal-01365953} with $5000<Re<25000$ performed lock-exchange experiments to report the mixing efficiency once the motion had completely ceased. Here, the density profiles were measured, before and after the end of the experiment, where the fluid had come to rest after reflection from the wall. The upper bound value of mixing efficiency was different in all the three separate studies but less than the generally accepted value of 0.17 used in oceanic flows (\citet{Osborn1980}). The reason for lower values of mixing efficiency could be attributed to the fact that it was integrated over the entire volume of the tank and the entire gravity current that also contained lesser energetic regions. Simultaneous Particle Image Velocimetry (PIV) and Planar Laser Induced Fluorescence (PLIF), which henceforth will be called simultaneous PIV-PLIF, gives an added advantage of capturing velocity and density data at discrete locations with unprecedented spatial and temporal resolutions. Using this technique, entrainment and mixing dynamics were reported by {\citet{odier_chen_ecke_2014}} for an inclined gravity current with continuous flow of fluid into an ambient with varying turbulence intensity. The structural development of the gravity current when it was laminar or turbulent was studied and with the help of turbulence statistics the mixing efficiency was reported. Recently, {\citet{doi:10.1063/1.5023033}} reported entrainment dynamics of a lock-exchange gravity current using simultaneous measurements of velocity and density field. They concluded that the flux entrainment coefficient undergoes a series of mixing transitions, depending on the mode of instability that governs the mixing between the two fluids. Despite plenty of literature on gravity currents, characterization of turbulence and mixing based on the flow energetics for a lock-exchange gravity current has not been reported. 

In this study, we focus on quantifying the terms in the turbulent kinetic energy budget equation, namely, turbulent kinetic energy ($K$), shear production ($P$), buoyancy flux ($B$) and dissipation rate ($\epsilon$) to understand the local dynamics of a lock-exchange gravity current when it is in the slumping phase. Based on the energetics, mixing efficiency and eddy diffusivities are then calculated to characterize the mixing. All the measurements are made exclusively the body of the gravity current. By using simultaneous PIV-PLIF, both large-scale and small-scale flow statistics are well-resolved, which otherwise is very difficult to measure. This also helps in resolving the two most common length scales in shear-stratified turbulence, namely, Kolmogorov ($L_{\kappa}$) and Ozmidov ($L_{O}$) length scales that form the backbone of the turbulence activity parameter or buoyancy Reynolds number, $Re_b$. When it comes to field scenarios, $Re$ loses its importance since it is difficult to measure. Therefore, for shear-stratified flows, $Re_b$ is a preferred parameter, which is relatively easier to measure in field campaigns and serves the purpose of dynamic similarity between experiments and field ({\citet{barry_ivey_winters_imberger_2001}}; {\citet{shih_koseff_ivey_ferziger_2005}}). The mixing efficiency and the eddy diffusivities are ultimately parameterized as a function of $Re_b$, which would serve as an useful input to the numerical models aimed at simulating shear-stratified flows. 

The paper is structured as follows: \S\ref{TKE} explains the budget equation, mixing efficiency and indirect forms of eddy diffusivities. In \S\ref{exp_fac} description of the experimental setup is given along with the averaging technique used for calculating the turbulence statistics. This is followed by results and discussion in \S\ref{ts_ls} $\&$ \S\ref{rif}. The main conclusions are given in \S\ref{conc}.

\section{Turbulent kinetic energy budget equation and mixing efficiency}\label{TKE}

The genesis of mixing and its energetics is attributed to turbulence in the system and thus making it important to look into the turbulent kinetic energy ($K$) budget equation ({\citet{kundu2015fluid}}).

\begin{widetext}
\begin{equation}\label{eqn:tke_eqn}
\frac{\partial K}{\partial t}+U_{j}\frac{\partial K}{\partial x_{j}}= \frac{\partial}{\partial x_{j}}\bigg(-\frac{1}{\rho_{0}} p'u'_{j} + 2\nu u'_{i}s'_{ij} - \frac{1}{2} u'^2_{i}u'_{j}\bigg) - 2\nu s'_{ij}s'_{ij} - u'_{i}u'_{j}\frac{\partial U_{i}}{\partial x_{j}} - \frac{g}{\rho_{0}} \rho'w'
\end{equation}
\end{widetext}
where, $K=\frac{1}{2}(u'^{2}+v'^{2}+w'^{2})$ is the turbulent kinetic energy, $u'$, $v'$, and $w'$ are the fluctuating components of the velocities in stream-wise ($x$), lateral ($y$), and vertical directions ($z$) respectively, $p'$ is the pressure fluctuation, $s'_{ij}$ is the fluctuating strain rate tensor, $\nu$ is the kinematic viscosity, and $g$ is the acceleration due to gravity. The total derivative on the left hand side is the rate of change of turbulent kinetic energy in a control volume in a Eulerian frame of reference. The temporal change is the unsteady component ($Un$) and is zero for stationary turbulence. The convective term represents spatial inhomogeneity and is zero when the turbulence is homogeneous. The first term on the right hand side is the divergence which is again zero for homogeneous turbulence. It is responsible for transport ($Tr$) of turbulent kinetic energy through turbulent pressure fluctuations, viscous diffusion and turbulent stress. Second term on the right hand side is the viscous dissipation of turbulent kinetic energy ($\epsilon$). Third term on the right hand side is the rate at which $K$ is produced through shear production ($P$) deriving its energy from the mean flow. The last term on the right is the buoyancy flux ($B$) or the rate at which $K$ is consumed in mixing. The $Un$ and $Tr$ terms in  (\ref{eqn:tke_eqn}) measure the rate of change of \textit{$K$} and the degree of homogeneity in the chosen control volume respectively. The only terms that have a definite exchange of energy and acts as a source or a sink are $P$, $B$, and $\epsilon$. 
Mixing efficiency, $Ri_{f}$ is defined as the fraction of the $K$ that is consumed during a mixing event which is used to bring  about irreversible increment in the background potential energy of the system encasing the mixing event. Under the assumptions of stationarity ($Un=0$) and homogeneity ($Tr=0$), the rate of change of $K$ and the divergence term in the chosen control volume is negligible and the mixing efficiency is defined as (\citet{venayagamoorthy_koseff_2016}): 

\begin{equation} \label{eqn:rif1}
Ri^I_{f}=\frac{B}{P} 
\end{equation}
\\
In (\ref{eqn:rif1}), the expression $\frac{B}{P}$ can also be written as ($\frac{B}{B+\epsilon}$) when the homogeneity and stationarity constraints prevail, which is seldom the case. Therefore, $Ri^I_{f}$ will predict the mixing incorrectly in non-homogeneous and non-stationary flows and one should account for the additional terms in the form of unsteadiness and inhomogeneity. This leads to the second definition of flux Richardson number, defined as $Ri^{II}_{f}$, first proposed by {\citet{doi:10.1175/1520-0485(1991)021<0650:OTNOTI>2.0.CO;2}}:

\begin{equation}
Ri^{II}_{f}=\frac{B}{-Un+Tr+P}=\frac{B}{B+\epsilon}
\end{equation}
The second definition of mixing efficiency ($Ri^{II}_{f}$) is generic and is widely applicable in shear-stratified flows. Rearranging the terms in (\ref{eqn:tke_eqn}), it is evident that $B$ and $\epsilon$ are enough to calculate $Ri^{II}_{f}$ and there is no need to calculate $Un$ and $Tr$ terms. 

The pitfall of using the above two definitions of $Ri_{f}$ is that it fails to segregate the down-gradient buoyancy flux (mixing) and the counter-gradient buoyancy flux (stirring). Down-gradient buoyancy flux brings about an irreversible change in the background potential energy of the system, whereas counter-gradient buoyancy flux is the reversible component that adds to the turbulent kinetic energy, $K$. For a mixing event to start, the eddies have to carry a heavier parcel of fluid into a lighter environment or vice-versa. This movement needs an external agent that is provided by $K$, which also decides to what extent the heavier and lighter fluid parcels mix. If $K$ is insufficient, the parcel does not lose its identity completely because of insufficient mixing and tries to revert to its new equilibrium position and this becomes the reversible components of the buoyancy flux ({\citet{doi:10.1146/annurev.fluid.35.101101.161144}}). It should be noted that the counter-gradient flux  (reversible component) cannot be eliminated completely in density stratified flows. However, it reduces as the turbulence activity increases and it has a more pronounced effect when the turbulence activity is low. Essentially, $K$ is the sole reason that promotes mixing and the intensity and the duration of it decides the dominance of the counter-gradient fluxes. There exists a third definition of mixing efficiency, $Ri_{f}^{*}$, that provides a fix for the effect of counter-gradient flux. The total energy of the system is compartmentalised into available and background potential energies according to a framework as explained by {\citet{winters_lombard_riley_d'asaro_1995}}. The $Ri_{f}^{*}$ uses a formulation based on the dissipation of available potential energy and $\epsilon$, that deals with the counter-gradient fluxes. To calculate $Ri_{f}^{*}$ from laboratory experiments is challenging and therefore is left as future scope. In the present study, we report only the first two definitions of mixing efficiency, which have not been reported in previous studies for lock-exchange gravity currents. A universal parameterization of $Ri_{f}$ is difficult because of the inherent variability in natural flows, which arises as a result of different mechanisms by which turbulent kinetic energy can be produced. This variability is evident from the discrepancies in the quantitative values of $Ri_{f}$ of field data at two different geographical locations and also from the laboratory experiments and a unified and an unambiguous framework therefore becomes difficult ({\citet{doi:10.1002/2014GL060571}}; {\citet{doi:10.1063/1.4868142}}). Therefore, it becomes interesting to delve into the energetics and mixing in a gravity current generated using a lock-exchange mechanism. The detailed analysis of the present problem can help in understanding many of the natural flows that have similar dynamics.

\begin{figure*}
  \centerline{\includegraphics[width=10cm, height=5cm]{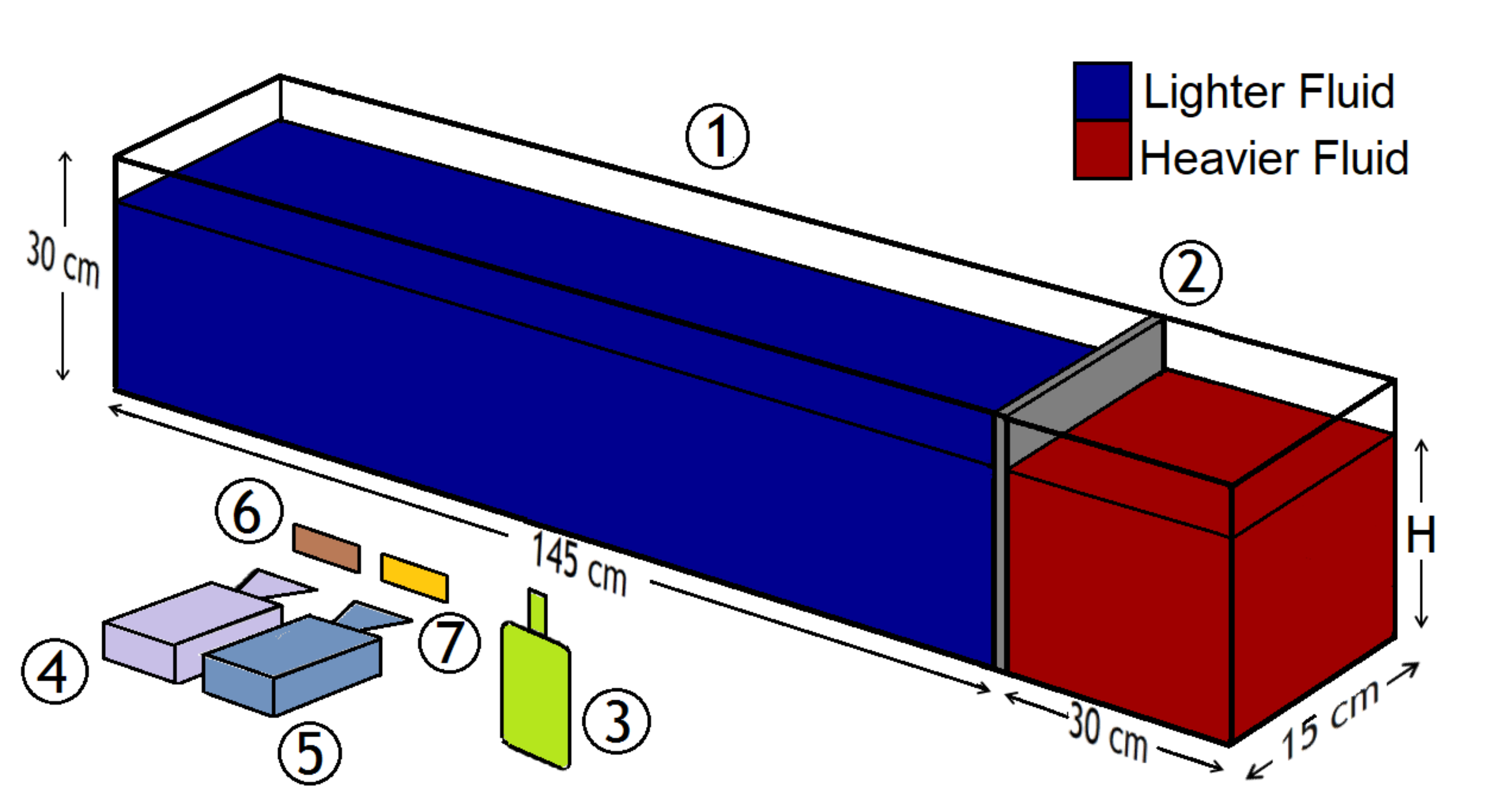}}
  \caption{Schematic of the experimental setup (not to scale). Isometric view of the setup, where 1 is the tank, 2 is the lock-exchange gate, 3 is the laser source illuminating the central portion of the tank, 4 and 5 are PIV and PLIF cameras, and 6, 7 are high-pass and low-pass filters.}
\label{fig:merged_setup}
\end{figure*}
\begin{figure*}
  \centerline{\includegraphics[width=14cm, height=3cm, trim = {0cm 8.5cm 0cm 0cm},clip]{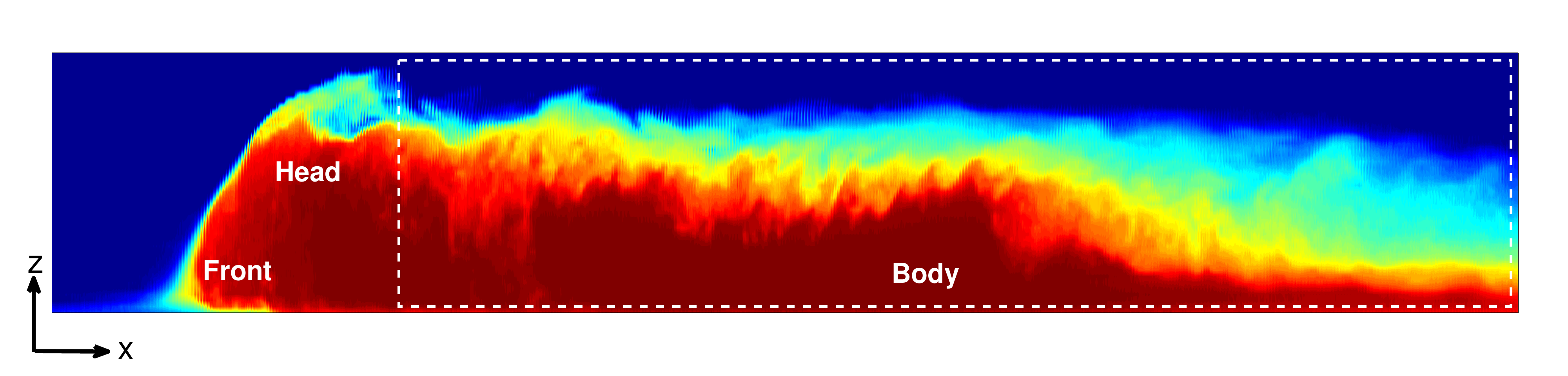}}
  \caption{Qualitative image of a lock-exchange gravity current, moving from right to left. The image is obtained by collating snapshots recorded at different time instances. Blue represents  ambient fluid (lighter) and red is the source fluid (denser). Intermediate colors are a result of mixing between the two fluids. Dashed white line shows the region where turbulence and mixing statistics are computed in the present study.}
\label{fig:naming}
\end{figure*}
The in-situ measurements of eddy diffusivities become difficult, as they are usually expressed in terms of turbulent fluctuating quantities that result from small-scale mixing. For a simple unidirectional flow, the eddy diffusivities are expressed as:

\begin{subequations}
\begin{equation}\label{eqn:km}
\kappa_{m}=-\frac{u'w'}{\frac{dU}{dz}}
\end{equation}

\begin{equation}\label{eqn:krho}
\kappa_{\rho}=-\frac{\rho'w'}{\frac{d\rho}{dz}}
\end{equation}
\end{subequations}
where, $u'w'$ is the Reynolds stress acting in the direction perpendicular to the flow to capture vertical mixing, $\rho'w'$ is the turbulent density flux, $\frac{dU}{dz}$ and $\frac{d\rho}{dz}$ are the mean vertical velocity and density gradients of the flow field. (\ref{eqn:km}) and (\ref{eqn:krho}) are the traditional ways of expressing eddy diffusivities, which has its pitfall since mathematically there exists a lack of closure because of Reynolds stress and density flux terms, moreover, simultaneous measurement of the fluctuating field variables are extremely challenging in field conditions. Therefore, one needs to resort to controlled laboratory experiments or numerical models that provide parameterizations for the eddy diffusivities of scalar and momentum. {\citet{Osborn1980}} proposed an indirect form of representation of eddy diffusivity of scalar as a function of mixing efficiency or flux Richardson number ($Ri_{f}$). A value of $Ri_{f}$ = 0.17 was proposed on the basis of controlled laboratory experiments conducted by { \citet{britter1974experiment}} and their field campaigns in the Atlantic ocean using stationarity and homogeneity conditions. We believe that $Ri_{f}$ from our present study would allow for a direct comparison with that reported by \citet{Osborn1980}, since in both these studies, the common sources of $K$ production are (a) Reynolds stress working against mean velocity gradient, and (b) collapsing Kelvin-Helmholtz billows (which are a typical feature of a lock-exchange gravity current). Using $Ri_{f}$, (\ref{eqn:km}) and (\ref{eqn:krho}) could be modified as,

\begin{subequations}
\begin{equation}\label{eqn:km_}
\kappa_{m}= \bigg(\frac{1}{1-Ri_{f}}\bigg) \frac{\epsilon}{S^2}
\end{equation}

\begin{equation}\label{eqn:krho_}
\kappa_{\rho}= \bigg(\frac{Ri_{f}}{1-Ri_{f}}\bigg) \frac{\epsilon}{N^2}
\end{equation}
\end{subequations}
where, $S$ is the mean shear rate (denoted as $S=d{U}/dz$) and $N$ (denoted as $\sqrt{-\frac{g}{\rho_0} \frac{\partial \rho}{\partial z}}$, where $\rho_0$ is a reference density and $\frac{\partial\rho}{\partial z}$ is the mean density field's gradient in the direction of acceleration due to gravity, $g$) is the buoyancy frequency of a parcel of fluid in a stably stratified environment or the Brunt-V$\ddot{a}$is$\ddot{a}$l$\ddot{a}$ frequency,. The way $Ri_{f}$  behaves over a domain of time and space gives insights about the dynamics of the flow and a parameterization of $Ri_{f}$  for shear-stratified flows is of paramount importance and a contemporary interest to the scientific community.

\section{Experimental set up and methodology}\label{exp_fac}

The experimental setup, shown in figure \ref{fig:merged_setup}, comprises of a plexiglass rectangular tank with a lock exchange gate provided to separate the two fluids initially. The tank dimensions are 175 cm long, 15 cm wide and 30 cm high. The tank is separated into two parts by installing a lock exchange gate 30 cm away from one of the ends, the smaller portion contains the heavier fluid (source) and the larger portion contains the lighter fluid (ambient). The heavier fluid is a salt-water solution and the lighter fluid is an ethanol-water based solution. This salt-ethanol based solution is employed so as to have matching refractive index of both the fluids. Details of the method of matching refractive index are available in {\citet{hannoun_fernando_list_1988}} and {\citet{doi:10.1007/s00348-012-1275-7}}. A densitometer and a refractometer are used to measure the density and match the refractive index of the two fluids respectively. The gate is lifted in one swift motion to reduce any secondary disturbances, resulting in a gravity current shown in figure \ref{fig:naming}. The swift motion ensures that the gravity current generated as a result of the density difference primarily controls the flow dynamics and any secondary disturbances bear minimal impact. A range of Reynolds numbers, varying from $Re$ = 485-12270, is realized by either varying the density of the two fluids or the height of the denser fluid in the initial lock-exchange position or both. Simultaneous measurements of velocity and density fields at different instances of time is obtained by employing simultaneous PIV-PLIF techniques, which is triggered using the same controller. The rate at which the data is collected is varied as per the flow velocity of the current. For low and intermediate $Re$, the cameras are set to collect the data at a rate of 50 fps and for high $Re$, it was set to 75 fps. The velocity field in $x$-$z$ (along the mean flow and perpendicular to it, in which the gravity acts) directions are obtained using PIV for a particular rectangular window located at the centre of the tank and the region is illuminated by a continuous-wave 500 mW, 532 nm laser from the bottom of the tank. The illuminated sheet measures 10 cm long and 10 cm wide and its center coincides with the center of the tank, which is $\approx$ 58 cm from the gate or $\approx$ 2 lock lengths away from the gate. This ensures that all the measurements were made during the slumping phase of the current. Neutrally buoyant particles made of polystyrene material of median diameter $\approx$ 15 $\mu$m are used as tracer particles to capture the flow field. Particle images are captured by IDS UI 3360 CP-M/C USB 3.0 camera with a resolution of 2048 by 1055 pixels and each pixel has an area of 5.5 $\mu m^2$. A lens is used with an aperture f/2 to reduce aberrations and to get an appropriate depth of field.

The density field is obtained using PLIF for the same window and using a separate camera. The laser source for both PIV and PLIF is the same. PLIF images are recorded by a IDS UI-1220 C USB 2.0 camera which has a CMOS sensor with 752 by 480 pixels. Rhodamine 6G (R6G) is used as a fluorescent dye that is mixed uniformly with the lighter fluid (only). The concentration of (R6G) in lighter and heavier fluid is 100 $\mu$g $L^{-1}$ and 0 $\mu$g $L^{-1}$ respectively. The gray value of the image and the R6G concentration is calibrated first, varying the concentration of R6G from 0 $\mu$g $L^{-1}$ to 100 $\mu$g $L^{-1}$ in increments of 10 $\mu$g $L^{-1}$. Decoding the image gray value, it is observed that R6G concentration is linearly proportional to the local density field. So, when the heavier (volume $V$ and density $\rho_{2}$) and the lighter fluid (volume $XV$ and density $\rho_{1}$, R6G concentration $C_{1}$) are mixed with each other, the concentration of R6G in that region changes as reflected by the images and the local density of that region can be found out:
$$\rho=\frac{\rho_{1}XV}{XV+V}=\frac{\rho_{1}X+\rho_{2}}{X+1}$$
$$C=\frac{C_{1}XV}{XV+V}=\frac{XC_{1}}{X+1}$$
Since the R6G concentration is known, the density at that location can be found using:

\begin{equation}
\rho=\rho_{2}-\frac{C}{C_{1}}(\rho_{2}-\rho_{1})    
\end{equation}
Therefore in the regions where the local concentration $C$ is equal to $C_{1}$, the local density $\rho$ is equal to $\rho_{1}$, i.e. the density of the lighter fluid. When the setup was in its initial position the local density field is same as that of the lighter fluid. However, once the gate is released and there is mixing between the two fluids, the local concentration of R6G changes signifying different local density.
 
Experiments with different values of Reynolds number ranging from $Re$ = 485-12270 are performed, as shown in Table \ref{tab:table1}. Note that all the experiments were a case of a full depth release, i.e. the initial depth of the heavier and the lighter fluids in the locked position are the same. Each set of experiment is conducted at least two times to ensure repeatability of the flow. The bulk parameters are chosen in a manner such that the $Re$ covers a spectrum of mixing regimes driven by waves and instabilities (\citet{doi:10.1063/1.5023033}). The $Re$ is calculated using $U_{F}$ and $h$, where, $U_{F}=0.4\sqrt{g^{'}H}$. Here, $h$ is defined as the length from the bottom of the gravity current to the first vertical location where the current density equals that of the ambient ($\rho_1$). For all the cases, $h$ $\approx$ $H/2$. The empirical constant in $U_{F}$ usually varies, most commonly between 0.4 to 0.5 depending on the fractional depth $\frac{h}{H}$ (Benjamin 1968). In our case, it is found to be 0.4. The heavier and the lighter fluids used in our experiments are completely miscible (that promotes mixing) and the flow is near inviscid in nature. 

\begin{table}[b]
\caption{\label{tab:table1}%
Experimental Parameters
}
\begin{ruledtabular}\label{tab:exp_par}
\begin{tabular}{ccccc}
\textrm{$\rho_1 (kg m^{-3})$}&
\textrm{$\rho_2 (kg m^{-3})$}&
\multicolumn{1}{c}{\textrm{Height, H (m)}}&
\textrm{$U_F (m s^{-1})$}&\textrm{Re}\\

\colrule
     
        996.1   & 1000.9 &   0.05 & 0.0194  & 485\\
        990.4   & 1009.9 &   0.05 & 0.0393  & 985\\
        980.4   & 1029.0 &   0.05 & 0.0623  & 1560\\
        990.3   & 1011.7 &   0.08 & 0.0521  & 2080\\
        993.2   & 1007.0 &   0.12 & 0.0511  & 3070\\
        980.4   & 1026.0 &   0.10 & 0.0854  & 4270\\
        971.6   & 1046.0 &   0.10 & 0.1096  & 5480\\
        980.3   & 1028.1 &   0.15 & 0.1071  & 8050\\
        971.4   & 1046.8 &   0.15 & 0.1351  & 10150\\
        971.2   & 1047.1 &   0.17 & 0.1443  & 12270\\

\end{tabular}
\end{ruledtabular}
\end{table}

For all the ten experiments, the data is gathered at the central region of the tank illuminated by a laser sheet. The 2-D PIV and PLIF measurement techniques have a vector resolution of 1.4 mm/vector and 0.35 mm/vector respectively. Since the PIV vector resolution is coarser, the data from PLIF is mapped on to the PIV grid to measure the mean and fluctuating components of velocity and density fields at the same spatial location, thereby enabling calculation of the scalar flux. It should be noted that the measurements are made once the highly transient head is past the illuminated area and all the results are restricted to the body of the current (as shown in figure \ref{fig:naming}). 

The mean field is calculated based on a moving average filtering method that is commonly used in field measurements (\citet{zhong_hussain_fernando_2018}). The filter width used in our study is $\Delta t U_F/H \approx 0.5$, that ensured the mean field is smooth and in concurrence with the trend expected for a lock-exchange current (i.e. decaying with time). The aim is to achieve a smooth mean field, such that further adjustments in the filter width will bear minimal impact on results. In general, the qualitative trends of the mean field and the turbulence statistics would be independent of the filter width, but the quantitative values may be a little sensitive to the choice of the filter width. A large number of images for the control volume (at least $>500$, by adjusting the frame rate of the camera) were used for all the $Re$ to reduce the statistical random error and for convergence (\citet{77f1936b918c405283750ff3fbe8b9d5}). If the evolution of any turbulence statistics, say, $X$ in the 2-D control volume has to found with time, it can be represented as:

\begin{equation} \label{eqn:avgt}
\overline{X}(t) = \frac{1}{n_x n_z}\int_{0}^{n_z} \int_{0}^{n_x} X dxdz
\end{equation}
where, $n_x$ and $n_z$ are the total number of finite grid points in the horizontal and vertical directions. Similarly, if the average variation of a quantity across the depth of the current has to be found, the following expression is used:

\begin{equation} \label{eqn:avgz}
\overline{X}(z) = \frac{1}{n_x n_t}\int_{0}^{n_x} \int_{0}^{n_t} X dtdx
\end{equation}
where, $n_t$ is the finite number of time frames or steps over which the quantity is averaged. Finally, if the quantity has to be averaged over time and for the entire control volume, the following was done:

\begin{equation} \label{eqn:avg}
\overline{X} = \frac{1}{n_x n_z n_t}\int_{0}^{n_z} \int_{0}^{n_x} \int_{0}^{n_t} X dtdxdz
\end{equation}
The gradient of the mean stream-wise velocity, $U$, in the lateral direction ($y$) is often discarded since the current spans the entire width. However, the fluctuations in the velocity in all the three directions will exist due to the nature of turbulence. The fluctuation in the $y$ direction is closely approximated to be equal to the fluctuation in $z$ direction (\citet{odier_chen_ecke_2014}). Hence, the turbulent kinetic energy $K$ is expressed as: 

\begin{equation}\label{eqn:k}
K=\frac{1}{2}(u'^{2}+ 2w'^{2})
\end{equation}
Since the gradient of the velocity exists primarily in the $z$ direction because of its unidirectional nature and $x$ direction because of the inhomogeneity in the body along its length, only two terms in the Reynolds stress tensor will contribute to the turbulent kinetic energy production, $P$ and is given as:

\begin{figure*}
  \centerline{\includegraphics[width=18cm, height=6cm, trim = {0cm 5cm 0cm 0cm},clip]{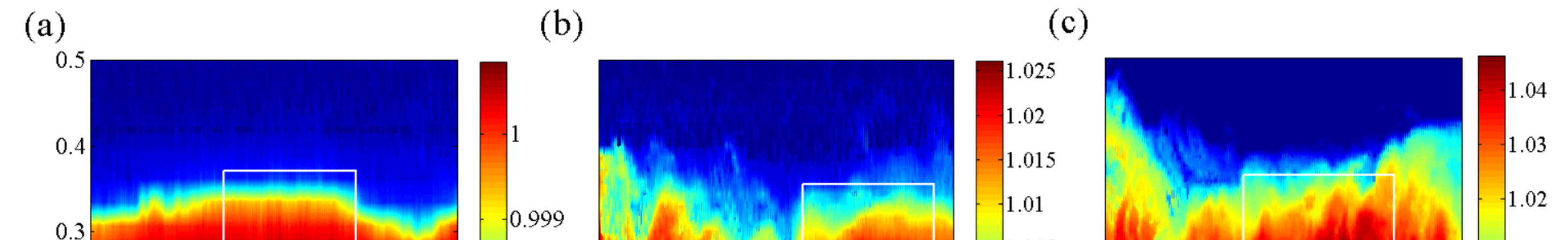}}
  \caption{Density contours sampled from the body of the current for (a) $Re$ = 485, (b) $Re$ = 4270, and (c) $Re$ = 12270 qualitatively capturing different energetic regimes. The figure also shows qualitatively, how the thickness of the mixing zone varies with $Re$. The colorbar indicates the densities across the height of the current in $g cm^{-3}$.}
\label{fig:density_contour}
\end{figure*}

\begin{equation}\label{p}
P=-u'w'\frac{dU}{dz}-u'u'\frac{dU}{dx}
\end{equation}
The buoyancy flux in the budget equation inhibits (or promotes) the turbulent kinetic energy production when the stratification is stable (or unstable) and is expressed as:

\begin{equation}\label{b}
B=\frac{g}{\rho_{0}}\rho'w'
\end{equation}
The estimation of turbulent kinetic energy dissipation rate, $\epsilon$, is an important aspect of the energetics of the flow that gives insights about the nature and degree of turbulence. Buoyancy Reynolds number, indirect form of the diffusivities, and turbulent length scales (Ozmidov scale and Kolmogorov scale), all  have a dependence on $\epsilon$. Two challenges that are imposed on precisely finding out the energy dissipation using PIV are: (a) the spatial resolution of PIV, it should be close to the smallest length scale in a turbulent flow (Kolmogorov length scale), otherwise $\epsilon$ would be severely underestimated and (b) For 2-D PIV, there exists an unresolved third component that has to be taken into account by making approximations using the other two resolved components of velocity. The smallest length scale is estimated as $L_{\kappa}=(\frac{\nu^3}{\epsilon})^{0.25}$, where $\epsilon \approx u'^3/h$ is a rough estimation. The smallest value of $L_{\kappa}$ corresponds to the highest Reynolds number, $Re$ = 12270, which was found to be $L_{\kappa}\approx$ 0.3 mm. The PIV resolution is 1.4 mm, giving the ratio of vector resolution ($\Delta$) to the Kolmogorov length-scale ($L_{\kappa}$) as $\frac{\Delta}{L_{\kappa}} \approx$ 4, which gives a fairly good estimation of $\epsilon$. Past studies on $\epsilon$ estimation by {\citet{doi:10.1021/ie0208265}}, {\citet{XU2013662}}, and {\citet{odier_chen_ecke_2014}} have revealed that when $\frac{\Delta}{L_{\kappa}} \leq$ 5, the dissipation in experiments can be resolved to a good extent within the statistical error limit. At low and moderate values of $Re$, the $\frac{\Delta}{L_{\kappa}}$  was found to be within $\frac{\Delta}{L_{\kappa}}\leq$ 3, which again is within the resolvable limit for $\epsilon$. The 2-D fluctuating strain rates (shear and normal) are computed using the values of fluctuating velocity components at every spatial location and the PIV resolution in the region of interest. The assumption of local isotropy ({\citet{tennekes1972first}}) is sometimes used as a simplification and estimates the dissipation as $\epsilon = 15\nu\big<{\big(\frac{du'}{dx'}}\big)^2\big>_{x,z}$, where only one of the gradients is required. However, this may over predict the dissipation in our case since the fluctuations in $y$ and $z$ directions are considerably lower than the fluctuations in the $x$ direction. In our experiments, PIV gives the entire 2-D velocity field and therefore we relax the assumption of isotropy and make use of all the available velocity gradients to estimate $\epsilon$ (see {\citet{doi:10.1175/1520-0485(2001)031<2108:TCADEI>2.0.CO;2}}) at every spatial location. The expression for $\epsilon$ takes the following form:

\begin{equation}\label{e}
\begin{aligned}
\epsilon=2\nu s'_{ij}s'_{ij}=\nu\bigg[4\bigg({\frac{\partial u'}{\partial x'}}\bigg)^2 + 4\bigg({\frac{\partial w'}{\partial z'}}\bigg)^2 \\+ 3\bigg({\frac{\partial u'}{\partial z'}}\bigg)^2 
+ 3\bigg({\frac{\partial w'}{\partial x'}}\bigg)^2 \\+ 4\frac{\partial u'}{\partial x'}\frac{\partial w'}{\partial z'} + 6\frac{\partial u'}{\partial z'}\frac{\partial w'}{\partial x'}\bigg]
\end{aligned}
\end{equation}
The above expression makes use of the incompressible continuity equation for fluctuating flow field to compute the fluctuating velocity gradient in the transverse (y) direction using the fluctuating velocity gradients in $x$ and $z$ directions ($\frac{\partial v'}{\partial y'}$ = -($\frac{\partial u'}{\partial x'}$+$\frac{\partial w'}{\partial z'}$)). Similarly, the unknown cross-fluctuating velocity gradient ($\frac{\partial u'}{\partial y'}$, $\frac{\partial w'}{\partial y'}$, $\frac{\partial v'}{\partial x'}$, $\frac{\partial v'}{\partial z'}$) is approximated as the average of the available fluctuating velocity gradients in $x$ and $z$ directions ($\frac{\partial u'}{\partial x'}$, $\frac{\partial u'}{\partial z'}$, $\frac{\partial w'}{\partial z'}$, $\frac{\partial w'}{\partial x'}$).

\section{Turbulence and mixing in gravity currents}\label{ts_ls}
\subsection{Modes of mixing and energetics}\label{ts}

\begin{figure*}
  \centerline{\includegraphics[width=14cm, height=5cm, trim = {0 3.5cm 0 0},clip]{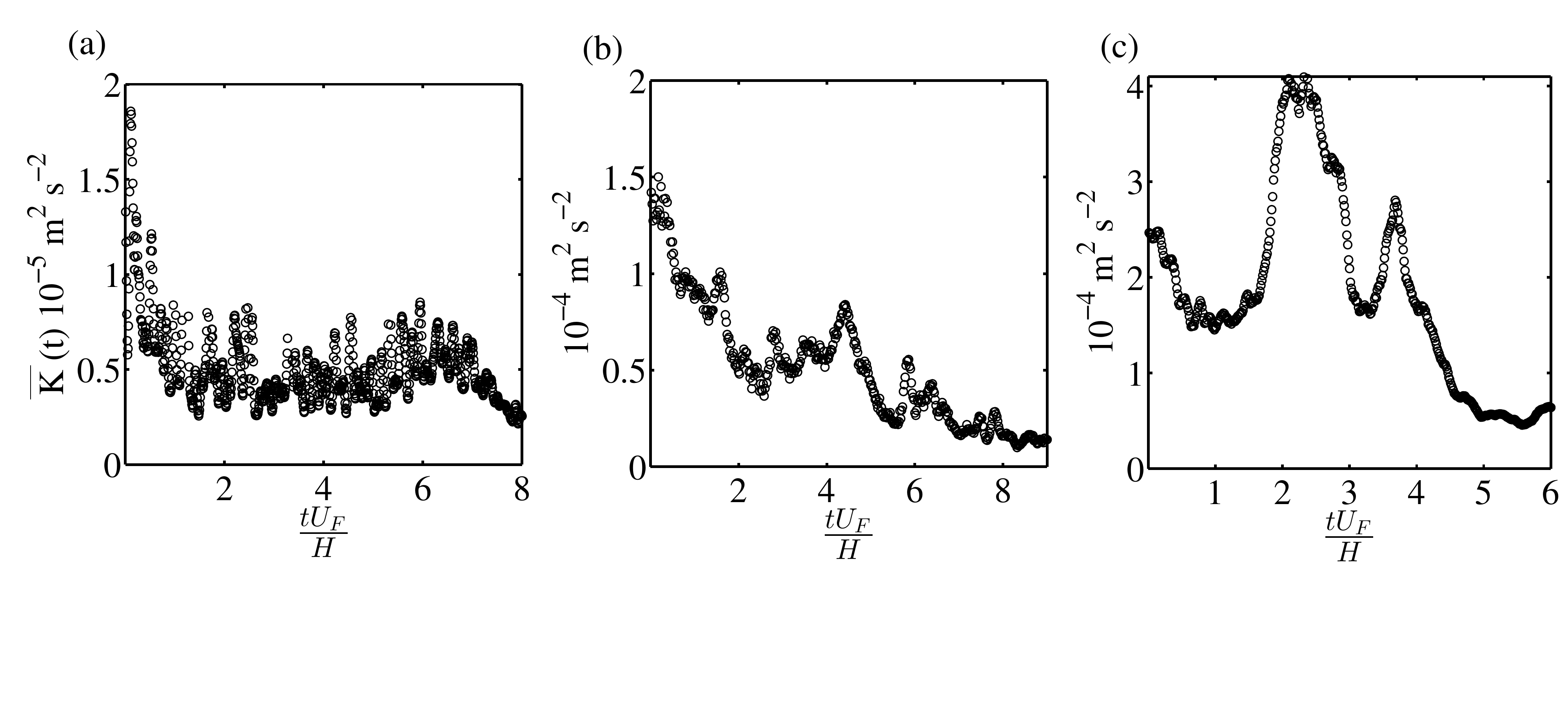}}
  \caption{Evolution of turbulent kinetic energy, $\overline{K}(t)$, in the body of the current for (a) $Re$ = 485, (b) $Re$ = 4270 and (c) $Re$ = 12270.}
\label{fig:tke_comb}
\end{figure*}

\begin{figure*}
  \centerline{\includegraphics[width=14cm, height=4.8cm, trim = {0 3.5cm 0 0},clip]{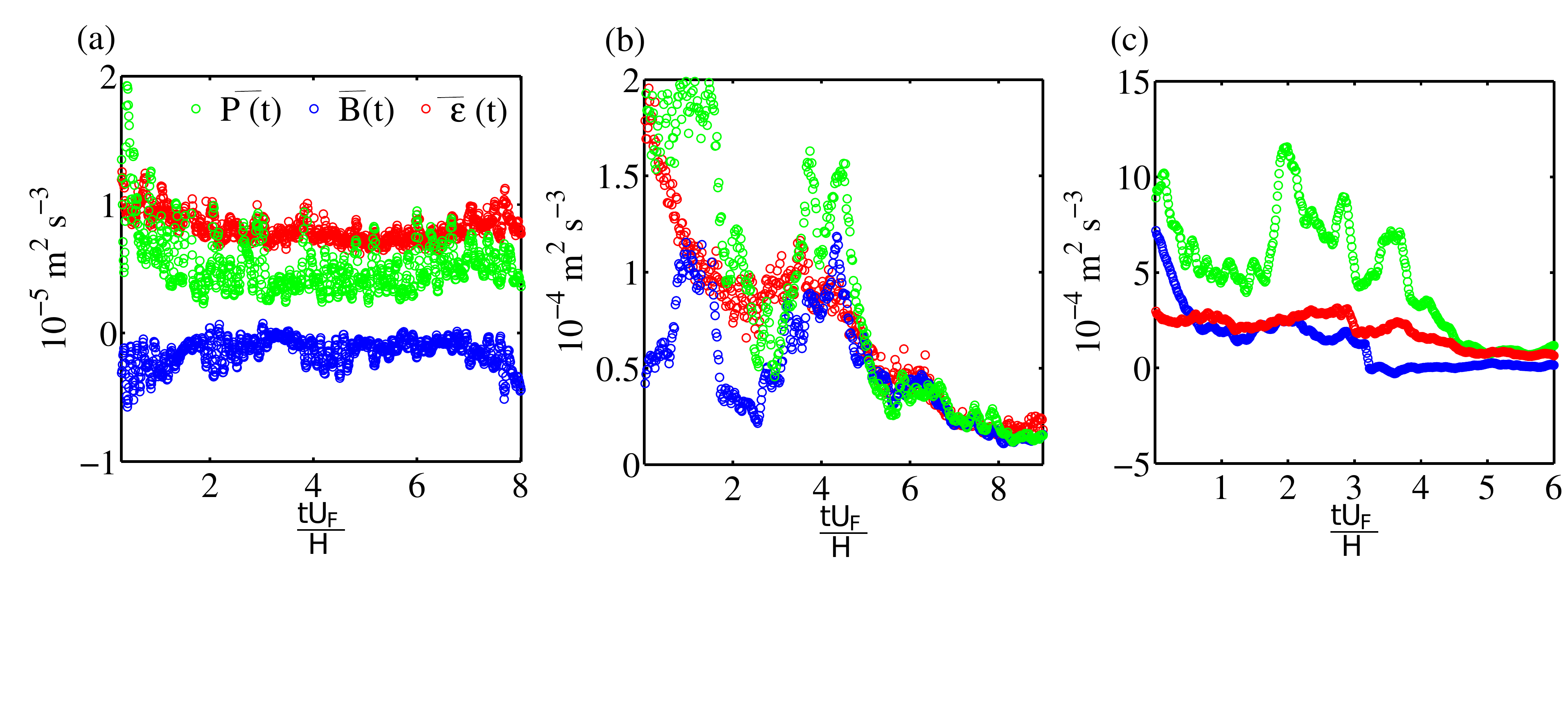}}
  \caption{Evolution of turbulence statistics $\overline{P}(t)$, $\overline{B}(t)$ and $\overline{\epsilon}(t)$ for (a) $Re$ = 485, (b) $Re$ = 4270 and (c) $Re$ = 12270.}
\label{fig:pbe_t}
\end{figure*}

Based on the thicknesses of velocity and density layers, {\citet{doi:10.1063/1.5023033}} demarcated different regimes of mixing, viz., Holmboe wave dominated mixing and Kelvin-Helmholtz instability dominated mixing (see figure \ref{fig:density_contour}). For a qualitative and a quantitative comparison, we consider three Reynolds numbers covering different mixing regimes, a low value, $Re$ = 485 (Holmboe waves), a moderate value, $Re$ = 4270 (Kelvin-Helmholtz instability), and the highest value, $Re$ = 12270 (energetic Kelvin-Helmholtz instability), and compare the turbulence statistics for a better understanding of the small-scale mixing dynamics. In figure \ref{fig:density_contour}, the density contour plots for the three values of $Re$ are shown. The depth of the current is non-dimensionalised by $H$. It can be seen that in the case of $Re$ = 485, the gravity current holds its shape and does not mix well with the ambient fluid. For low values of $Re$ ranging from $Re$ = 485-1560, the instabilities are weak and turbulence and mixing are primarily driven by Holmboe waves as shown in figure \ref{fig:density_contour}(a). The time evolution of turbulent kinetic energy, $\overline{K}(t)$, for the lowest $Re$ case is shown in figure \ref{fig:tke_comb}(a). It is observed that $\overline{K}(t)$ rapidly reduces upto $\frac{tU_{F}}{H}\approx$ 2 and later attains a near constant value. This indicates that the turbulence decays quickly due to less energy present in the flow at low values of $Re$. A mixing transition is observed after $Re$ = 2080, where the turbulence and associated mixing is now controlled by Kelvin-Helmholtz instability giving rise to vortical structures (see $Re$ = 4270 in figure \ref{fig:density_contour}(b)). A gradual decay in $\overline{K}(t)$ is observed over the entire time period. This possibly indicates presence of large and small-scale structures in the flow that transfer energy between themselves, thereby resulting in a monotonic decay of turbulence. Beyond $Re$ = 8000, a highly turbulent state is achieved and an even more energetic Kelvin-Helmholtz instability is observed. Large vortical structures in the shape of billows are formed that eventually break down resulting in efficient mixing (see figure \ref{fig:density_contour}(c)). The effect of the turbulence near the interface has a deeper penetration in the unstratified bulk of the gravity current present at the bottom and the motion is quite chaotic. Unlike $Re$ = 485, the mixing at higher values of $Re$ is predominantly dictated by turbulent diffusion, leaving a trail of thoroughly mixed fluid in its wake. For $Re$ = 12270, the unstratified bulk (in dark red) is almost being ``eroded" which is absent in lower $Re$ cases. The peak in the value of $\overline{K}(t)$ occurring at a later time ($\frac{tU_{F}}{H}\approx$ 2) for $Re$ = 12270 may be attributed to episodic bursts of energy due to the collapsing K-H billows that may increase $\overline{K}(t)$ momentarily.

\begin{figure*}
  \centerline{\includegraphics[width=14cm, height=4.5cm, trim = {0 5.8cm 0 0},clip]{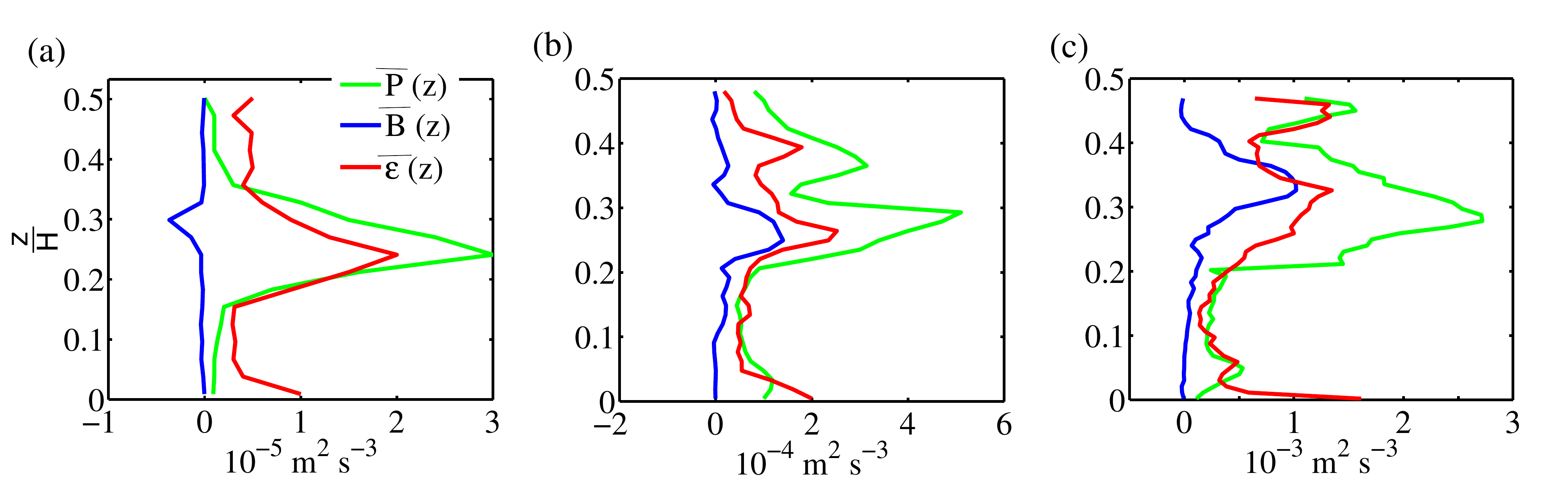}}
  \caption{Average turbulence statistics, $\overline{P}(z)$, $\overline{B}(z)$, and $\overline{\epsilon}(z)$ across the depth of the current (a) $Re$ = 485, (b) $Re$ = 4270 and (c) $Re$ = 12270.}
\label{fig:p_b_e}
\end{figure*}

\begin{figure*}
  \centerline{\includegraphics[width=14cm, height=5.3cm, trim = {0 3.5cm 0 0},clip]{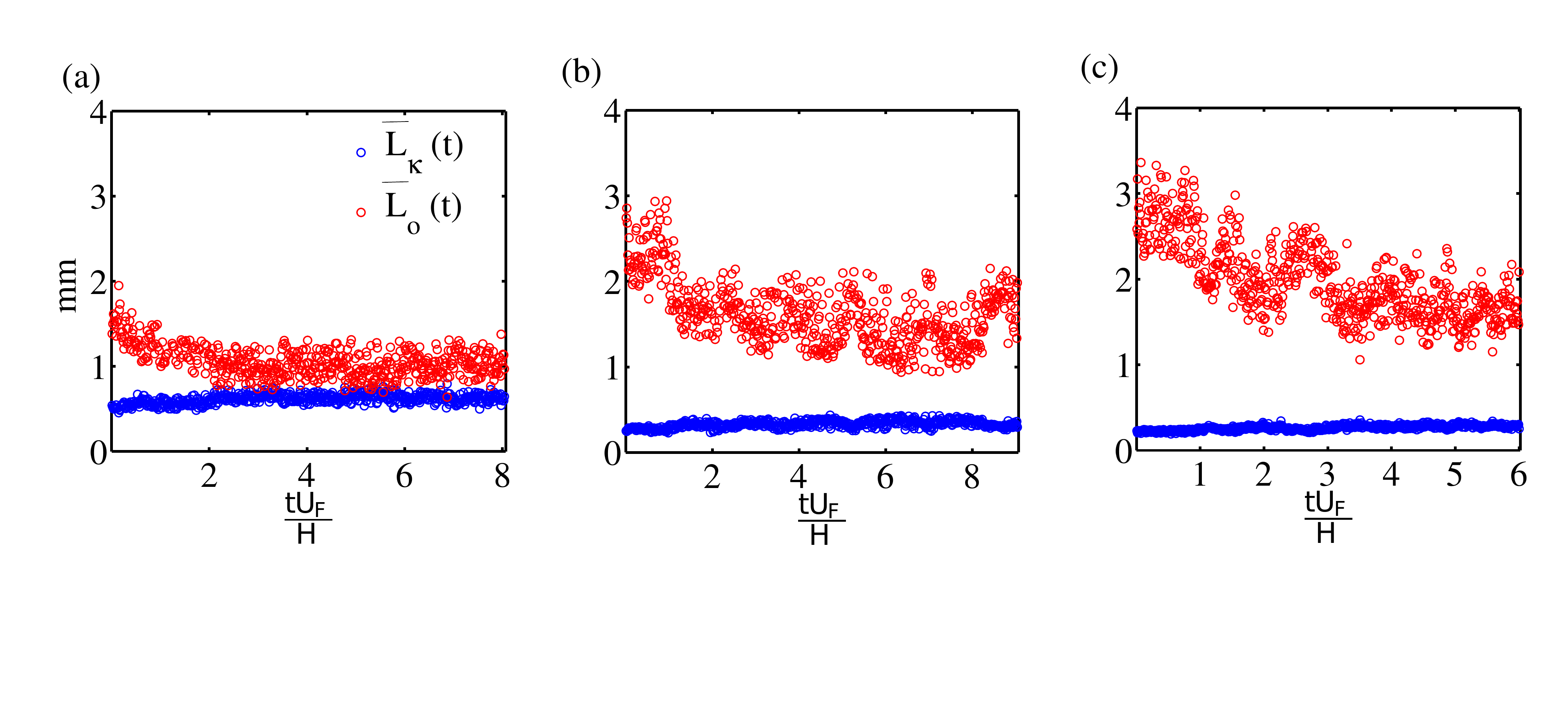}}
  \caption{Evolution of Ozmidov scale, $\overline{L_{o}}(t)$, and Kolmogorov scale, $\overline{L_{\kappa}}(t)$, for (a) $Re$ = 485, (b) $Re$ = 4270 and (c) $Re$ = 12270.}
\label{fig:length_scales}
\end{figure*}

To complete the discussion related to the energetics within the gravity current, the other terms in the budget equation, which are the source and sink of $K$ also has to be found [see eqns (\ref{p}), (\ref{b}), (\ref{e})].  A closure or a balance is required to understand the energetics of a shear-stratified flow and to understand the different ways in which the energy is generated and distributed. Using (\ref{eqn:avgt}), the evolution in the shear production rate, the buoyancy flux and the viscous dissipation rate are found with time (see figure \ref{fig:pbe_t}). The quantitative and qualitative behaviour of these quantities are quite similar to that of the $K$'s evolution with time, $\overline{K}(t)$, but their competing nature is more interesting to look at.A balance between $\overline{P} \approx \overline{B} + \overline{\epsilon}$ means that the system is homogenous and stationary and this happens rarely. In our case, for $Re =$ 485, this balance is almost intact for the entire time period. Similarly, for $Re =$4270, 12270 a balance is seen at a much later stage, when the energy of the current becomes quite low. In figure \ref{fig:p_b_e}, vertical variations of $\overline{P}(z)$, $\overline{B}(z)$, and $\overline{\epsilon}(z)$ are presented. It could be seen that most of the turbulence activity and energetics are confined to the region close to the interface of the gravity current fluid and the ambient fluid, i.e., within the shear layer. The vertical axis in figure \ref{fig:p_b_e} is non-dimensionalized with $H$ to highlight the relative thickness of the mixed layer compared to the depth of the current. The vertical extent of the cusp formed by the buoyancy flux $\overline{B}(z)$ in figure \ref{fig:p_b_e} gives an estimate of the mixing zone and it becomes deeper as the $Re$ increases.  A significant change in the energetics is observed  near the interface due to the presence of interfacial Holmboe waves or Kelvin-Helmholtz type instabilities that result in an overturning moment, which allows the gravity current to engulf the ambient fluid through the swirling motion of the eddies. This engulfing action, by large, is absent in low $Re$ cases and the mixing is primarily due to weak interfacial waves. This leads to the existence of counter-gradient fluxes dominating the flow that can be noticed for $Re$ = 485 from the buoyancy flux plot in figure \ref{fig:p_b_e}(a), wherein the value of $\overline{B}(z)$ is negative in the mixing zone. At $Re$ = 485, the turbulence activity is so small that the irreversible component of buoyancy flux is masked by its reversible component and therefore shows negative values in the mixing zone. On the other hand, the buoyancy flux produced in the regime of moderate and high $Re$ is predominantly down-gradient, resulting in a net positive value of it, signifying that a part of the turbulent kinetic energy of the flow is converted to the background potential energy of the system. The relative magnitudes of dissipation rate, $\overline{\epsilon}(z)$ and shear production rate, $\overline{P}(z)$ provide us with some valuable information about the flow. Usually both these quantities are of the same magnitude in the absence of any stratification, suggesting that the energy dissipates only into heat. However, when there is a density stratification present, it results in a positive buoyancy flux that uses a part of the turbulence kinetic energy. For low $Re$ case, even in the presence of stratification, the magnitudes of shear production rate and dissipation rate are almost of similar magnitudes (see figure \ref{fig:pbe_t}(a) and \ref{fig:p_b_e}(a)) which indicates an presence of a very weak turbulent state. Another feature worth noting in the dissipation characteristics that is common amongst all the cases is the secondary dissipation peak that is formed at the bottom (see figure \ref{fig:p_b_e}). This is because of the bottom surface of the tank or the bed acts as a solid wall and prevents the adjacent fluid layer to move further because of no-slip condition creating a secondary bottom shear layer. In essence, the turbulence activity is limited to a region near the interface where all the turbulence statistics show a peak, the instabilities are a result of relative velocities between the strata of fluids with different densities that give rise to a classic case of Kelvin-Helmholtz instability (and Holmboe waves for low $Re$ case).
  
\subsection{Length-scales in shear-stratified turbulence}\label{reb}

Another important metric in a shear-stratified flow is the turbulent length scales. Two widely used length scales for the characterization of shear-stratified flows are the Kolmogorov ($L_{\kappa}$) scale and the Ozmidov ($L_{o}$) scale. The Kolmogorov length scale ($L_{\kappa}$) is the smallest length scale at which energy contained in the larger eddies dissipates into heat. On the other hand Ozmidov length scale ($L_{o}$) is the smallest length scale that can be influenced by the stratification. The stratification preferentially deforms the larger scales first and then the smaller scales, therefore Ozmidov scale is the size of the smallest eddy that can be deformed and whose energy can be suppressed by the stratification strength. These two length scales form the backbone of turbulence activity parameter or buoyancy Reynolds number ($Re_{b}$), which is the ratio of Ozmidov length scale to the Kolmogorov length scale. The length scales and the buoyancy Reynolds number are expressed as:

$$\overline{L_{o}}(t)=\bigg(\frac{\overline{\epsilon}(t)}{N^3}\bigg)^{\frac{1}{3}} \quad  \quad  \overline{L_{\kappa}}(t)=\bigg(\frac{\nu^3}{\overline{\epsilon} (t)}\bigg)^{\frac{1}{4}} \quad$$

\begin{equation}
Re_{b}=\bigg(\frac{L_{o}}{L_{\kappa}}\bigg)^{\frac{4}{3}}=\frac{\overline{\epsilon}}{\nu N^2}
\end{equation}
The average change in these length scales for the three different $Re$ values are shown in figure \ref{fig:length_scales}, which gives more quantitative information about the state of turbulence. A general trend that is seen in all the plots is that the Kolmogorov length scale increases with the stream-wise direction, $x$, while the Ozmidov length scale decreases. In general for all the three $Re$ values, it can be seen that $\overline{L_{o}}(t)$ is the highest just behind the head, where the body begins, indicating that the buoyancy or stratification only affects the large-scale eddies. As we move downstream, $\overline{L_{o}}(t)$ decreases due to reduced inertial forces and the effect of stratification is now felt on smaller eddies as well. In contrast, $\overline{L_{\kappa}}(t)$ is smallest at the initial times when the head has just flushed out, indicating high turbulence activity just behind the head and the initial parts of the body. As we move downstream, the value of $\overline{L_{\kappa}}(t)$ increases, signifying decaying turbulence and reduced mixing (as stratification suppresses turbulence production). It can be seen that for $Re$ = 485, the length scales do not evolve significantly with time, indicating that the level of turbulence activity is almost the same in the horizontal span of the particular region. This is consistent with our earlier argument that the turbulence activity is quite low and the current did not attain a turbulent state. In the more energetic regimes, namely $Re$ = 4270 and $Re$ = 12270, there is a clear evolution of $\overline{L_{o}}(t)$ and $\overline{L_{\kappa}}(t)$ with time. There is a gradual decrease in the values of $\overline{L_{o}}(t)$ accompanied by a gradual increase in the values of $\overline{L_{\kappa}}(t)$, which indicates that the turbulent activity continuously decreases as we move downstream of the current's body. This signifies that the amplitude of fluctuations in field variables about its mean quantity constantly reduces as we moved downstream resulting in reduced turbulence intensity. The decrease in $\overline{L_{o}}(t)$ and the increase in $\overline{L_{\kappa}}(t)$ show that the rate at which the turbulence activity decays is directly proportional to the intensity of turbulence itself ($-\frac{d\overline{K}(t)}{dt} \propto \overline{K}(t)$), i.e. larger the turbulence activity, stronger is the decay rate, provided there is no external force energizing it. 

\begin{figure*}
  \centerline{\includegraphics[width=14cm, height=6.2cm, trim = {0 0.9cm 0 0},clip]{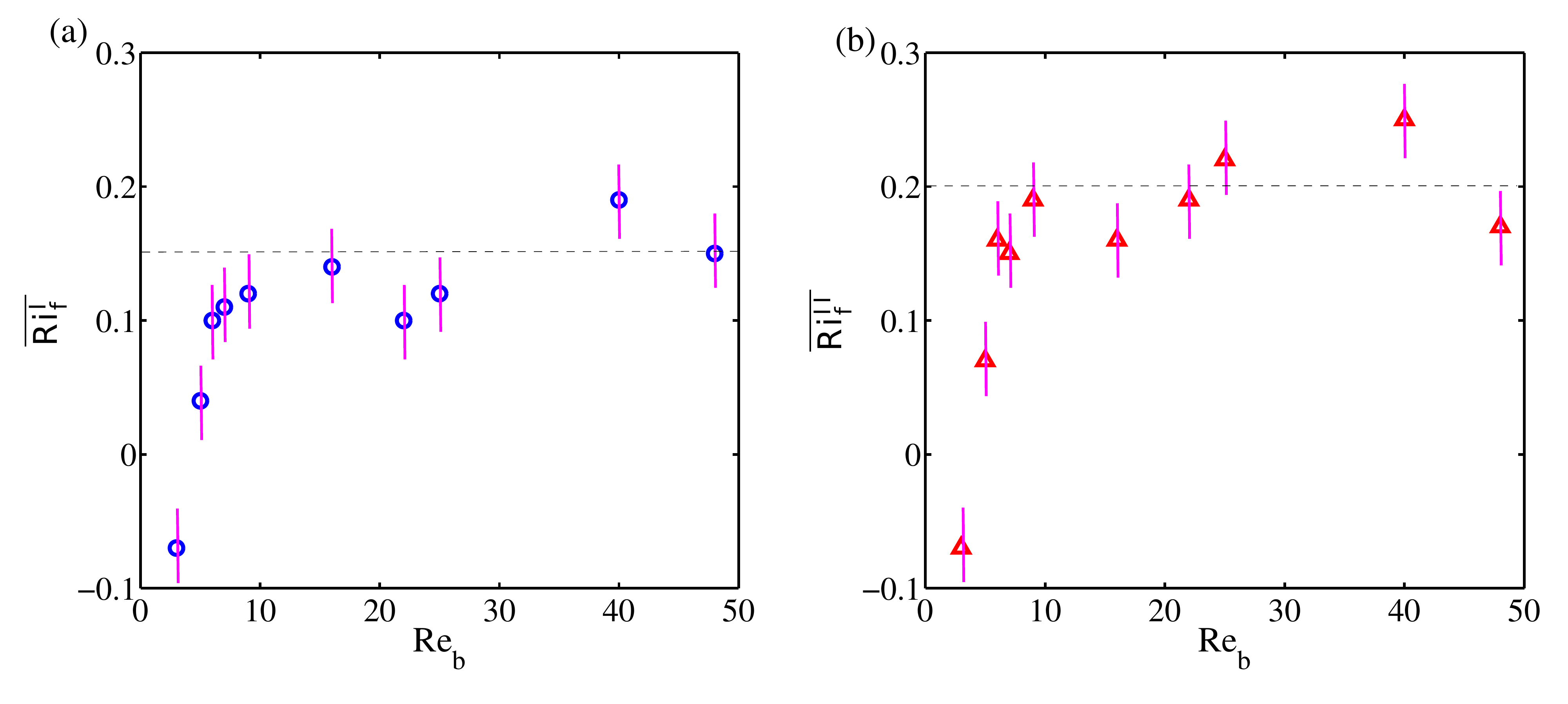}}
  \caption{Mixing efficiency ($\overline{Ri_f^I}$ and $\overline{Ri_f^{II}}$) as a function of buoyancy Reynolds number, $Re_b$. The magenta lines account for all the uncertainties that could arise in the calculations.}\label{fig:rif_reb}
\end{figure*}

\section{Mixing efficiency and eddy diffusivities of momentum and scalar}\label{rif}

The evaluation of mixing efficiency, $\overline{Ri_{f}}$, based on the flux terms in the turbulent kinetic energy budget equation is reported in this section. The mixing efficiency calculation helps in parameterizing the eddy diffusivities in shear-stratified flows, since field measurement of these diffusivities is challenging. As already discussed, the calculation of $\overline{Ri_{f}}$ requires simultaneous measurements of buoyancy flux and shear production rate, which yields the first definition of the mixing efficiency, viz. $\overline{Ri_{f}^I}$. By measuring the viscous dissipation in the budget equation, we  arrive at the second definition of mixing efficiency, $\overline{Ri_{f}^{II}}$, which is free from the homogeneous and stationary assumptions. Here, we report the quantitative values of both $\overline{Ri_f^{I}} = \frac{\overline{B}}{\overline{P}}$ and $\overline{Ri_f^{II}} = \frac{\overline{B}}{\overline{B}+\overline{\epsilon}}$ as a function of $Re_{b}$ in figure \ref{fig:rif_reb}, by making use of (\ref{eqn:avg}). 

\begin{figure*}
  \centerline{\includegraphics[width=14cm, height=6.5cm, trim = {0 0cm 0 0},clip]{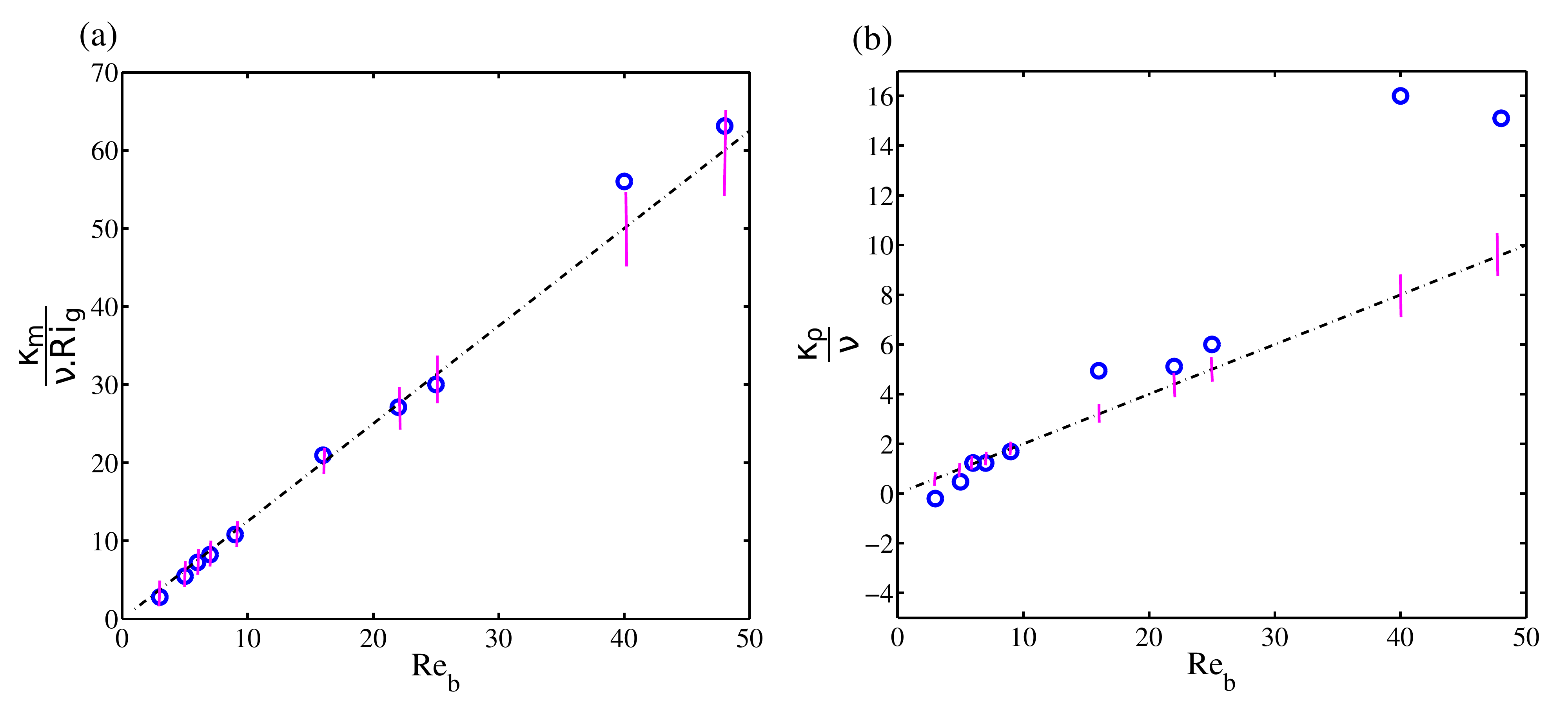}}
  \caption{(a) Normalized eddy diffusivity of momentum ($\kappa_m$) and the dashed line (-{}-{}-) represents 1.25 $Re_b$ (Crawford 1982). (b) Normalized eddy diffusivity of scalar ($\kappa_{\rho}$) and the dashed line (-{}-{}-) represents 0.2 $Re_b$ (Osborn 1980). The eddy diffusivities are parameterized using the second definition of flux Richardson number, $\overline{Ri_{f}^{II}}$. The magenta line shows deviation ($\pm$ 10 $\%$) from Crawford's and Osborn's parameterization.}
\label{fig:eddy_reb}
\end{figure*}

The $Re_{b}$ in our case ranged from 3 $<Re_{b}<$ 48, that covers the entire diffusive regime and a part of the intermediate regime (for details refer to {\citet{shih_koseff_ivey_ferziger_2005}}), which is very much relevant in oceanic flows. It should be noted here that the multiple values of $\overline{Ri_f^{I}}$ and $\overline{Ri_f^{II}}$ for different runs of $Re$  are averaged, and a single value for a particular $Re_b$ is reported. From figure \ref{fig:rif_reb}, it can be seen that for $0 < Re_b <10$, $\overline{Ri_f}$ starts from a low value and monotonically increases. The low mixing efficiency is attributed to low irreverisble buoyancy flux production, since most of the turbulent kinetic energy is expended in de-stabilizing a bottom heavy system (highly stable), which is also evident from figures \ref{fig:density_contour}(a) \& \ref{fig:p_b_e}(a). In this range of $Re_b$, the effects of counter-gradient (or up-gradient) fluxes are prominent, and its relative magnitude is higher in comparison to the down-gradient or irreversible buoyancy fluxes. In the intermediate regime ($Re_b>$ 10), it is seen that the mixing efficiency, $\overline{Ri_f}$, very nearly plateaus. As $Re$ increases, the turbulent kinetic energy increases and it promotes higher degree of small-scale mixing, which aids in increasing the magnitude of the down-gradient buoyancy fluxes. Therefore, in the more energetic regime, $\overline{Ri_f}$ increases and more importantly the effect of counter-gradient flux is less compared to the low $Re_b$ cases. Both $\overline{Ri_f^{I}}$ and $\overline{Ri_f^{II}}$ are plotted and juxtaposed together to notice the differences between the two definitions of mixing efficiency. Though the individual quantitative values of the turbulence statistics increase with an increase in $Re$ or $Re_b$ (see figure \ref{fig:p_b_e}), both the definitions of $Ri_{f}$ plateau at higher values of $Re_b$ (within the experimental parameter range) because of the competing effects of all the terms in the budget equation. The values of $\overline{Ri_f^{II}}$ are slightly higher than the values of $\overline{Ri_f^{I}}$, which implies that the assumption of homogeneity and stationarity in the body of the current does not hold. It should be pointed out that the second definition consistently predicts a higher mixing efficiency, since, our experiments revealed that $\overline{P}-\overline{B}-\overline{\epsilon}$ is positive for almost all values of $Re$ (except $Re$ = 485). Using (\ref{eqn:tke_eqn}), we can estimate the contribution of the transport term ($Tr$). Due to the decaying nature of turbulence in our system, the unsteady term ($Un$) in (\ref{eqn:tke_eqn}) is negative. Rewriting (\ref{eqn:tke_eqn}) as \textit{B+$\epsilon$=P+Tr-Un}, we note that in order to satisfy our experimental claim that $\overline{P}-\overline{B}-\overline{\epsilon}$ is positive, $Tr$ has to be negative. This means that part of the $K$ is used in maintaining the homogeneity, thereby inducing higher values of $\overline{Ri_f^{II}}$. The magenta lines in the figure \ref{fig:rif_reb} are provided to account for all the possible experimental uncertainties that can arise which may affect the calculation of fluxes. The values of mixing efficiency and the relative difference between the two definitions show dependence on the kind of $K$ production mechanism and the turbulence regime. At very low $Re$ ($Re<$\textit{500}) the two definitions are the same and collapse at a single point ($\overline{Ri_f^{I}}$= $\overline{Ri_f^{II}}$= -0.07, for $Re$ = 485), implying that homogeneity and stationarity conditions can be achieved at a weak turbulent state for this particular genre of flows, which is also consistent with our earlier discussions in \S\ref{ts_ls}. The values of mixing efficiency from figure \ref{fig:rif_reb} indicate that $\overline{Ri_f^{I}}$ saturates to $\approx$ 0.15 and $\overline{Ri_f^{II}}$ to $\approx$ 0.2 with a variability (which are inherent and inevitable in these flows) of $\pm 0.05$ that will envelope all the mixing efficiency values in that region ($Re_b>10$). Both these values, calculated from experiments, deviate slightly from the value of 0.17 proposed by Osborn, but are well within the statistical error limit.

The mixing efficiency, $\overline{Ri_{f}}$, helps in parameterization of eddy diffusivity of momentum and scalar as a function of buoyancy Reynolds number, $Re_{b}$. The ultimate aim is to be in a position to specify the effect of turbulence in enhancing mixing at molecular levels and it is quantified using the turbulent diffusivities. One way is to use eddy viscosity hypothesis (refer  (\ref{eqn:km}) and (\ref{eqn:krho})), but it often becomes difficult to measure the diffusivities directly in that manner. Therefore the eddy diffusivities are parameterized in the form of $\overline{Ri_f}$ and flow parameters that can be measured easily. The second definition of flux Richardson number ($\overline{Ri_f^{II}}$), which is more generic and does not use approximations, is used to parameterize the eddy diffusivities in our case. Figure \ref{fig:eddy_reb}(a), (b) show the normalized eddy diffusivity of momentum and scalar respectively as a function of $Re_{b}$. The normalization follows \citet{Osborn1980} to express eddy diffusivities as a function of measurable field variable, i.e., $Re_b = \frac{\overline{\epsilon}}{\nu N^{2}}$.

$$\kappa_{m}= \bigg(\frac{1}{1-\overline{Ri_{f}}}\bigg) \frac{\overline{\epsilon}}{S^2}=\bigg(\frac{1}{1-\overline{Ri_{f}}}\bigg)Ri_{g} \frac{\overline{\epsilon}}{N^2}$$

\begin{equation}\label{eqn:km_par}
\frac{\kappa_{m}}{\nu.Ri_g}= \bigg(\frac{1}{1-\overline{Ri_{f}}}\bigg) \frac{\overline{\epsilon}}{\nu N^2}
\end{equation}

$$\kappa_{\rho}= \bigg(\frac{\overline{Ri_{f}}}{1-\overline{Ri_{f}}}\bigg) \frac{\overline{\epsilon}}{N^2}$$

\begin{equation}\label{eqn:krho_par}
\frac{\kappa_{\rho}}{\nu}= \bigg(\frac{\overline{Ri_f}}{1-\overline{Ri_{f}}}\bigg) \frac{\overline{\epsilon}}{\nu N^2}
\end{equation}
With a limited number of data points that were available through  our experiments on lock-exchange gravity current and using the parameterization given in (\ref{eqn:km_par}) and (\ref{eqn:krho_par}), it can be seen from figure \ref{fig:eddy_reb}(a) that the eddy diffusivity of momentum is within $\pm$ 10 $\%$ of the parameterized form proposed earlier by \citet{crawford}. Also, from figure \ref{fig:eddy_reb}(b) it is evident that the values of eddy diffusivity of scalar are in close agreement within $\pm$ 10 $\%$ with the parameterization proposed by \citet{Osborn1980}, but only for a very limited range, $Re_{b}<$ 10, or the diffusive regime ({\citet{shih_koseff_ivey_ferziger_2005}}). In this region, for $Re_b<5$, the total diffusivity is less than zero, a result that is attributed to the presence of the counter-gradient fluxes. The negative values do not make sense in a practical scenario, since the third definition of mixing efficiency has not been resolved, that probably should act as a remedy to this problem, which we are currently working on. For now, in this study, negative values simply mean that the turbulence is weak and the turbulent diffusivity has not enhanced molecular diffusion appreciably for low $Re_b$ cases. Beyond $Re_b>10$ we see a marked departure indicating that the scalar eddy diffusivity behaves differently compared to the parameterized form proposed by {\citet{Osborn1980}} in this regime. Compared to the experimental results, the Osborn's parameterization under-predicts the eddy diffusivity in this regime and differ by a factor of about two, which is indicative of the dependence on the mode of mixing. Therefore, a different parameterization is required for modeling the scalar eddy diffusivity in this regime, underscoring the importance of this study.

\section{Conclusions}\label{conc}

The energetics and mixing efficiency in the body of a lock-exchange gravity current, while in its slumping phase, for varying Reynolds numbers ($Re$) and corresponding buoyancy Reynolds numbers ($Re_{b}$) have been quantified. The important terms in the turbulent kinetic energy budget equation, such as the turbulent kinetic energy ($K$), shear production ($P$), buoyancy flux ($B$), and energy dissipation ($\epsilon$) were calculated based on the simultaneous measurements of velocity and density fields using Particle Image Velocimetry (PIV) and Planar Laser Induced Fluorescence (PLIF) techniques respectively. The turbulence statistics indicated that most of the mixing activity in the body of the gravity current was restricted close to the interface of the ambient fluid and the gravity current and not the entire depth of it. The extent of the mixing zone relative to the current depth was found to increase with increasing Reynolds number owing to transition from weak Holmboe waves to Kelvin-Helmholtz instability. Moreover, it was observed that for low values of $Re$, the turbulence statistics, namely, $\overline{K}(t)$, $\overline{P}(t)$, $\overline{B}(t)$, $\overline{\epsilon}(t)$, decay rapidly indicating low level of turbulence. A more monotonic decay is observed at higher $Re$ values owing to presence of large-scale and small-scale flow structures that transfer energy amongst themselves. The values of $\overline{P}(z)$, $\overline{B}(z)$, and $\overline{\epsilon}(z)$ show a marked increase at the interface of the current due to the presence of a shear-driven mixed layer.

The two important length scales that rely heavily on dissipation, showed a trend that implied that the turbulence in the particular region decayed as we moved downstream, and the decay rate was proportional to the turbulence intensity. In order to quantify the local mixing dynamics, the average mixing efficiency, $\overline{Ri_f}$, was calculated using the turbulent flux terms. Two different representations of $\overline{Ri_f}$ were used. After a critical value of $Re>2080$ and $Re_{b}>10$, the mixing efficiency value was found to have an upper bound of $\overline{Ri_f^{I}}\approx 0.15$ and $\overline{Ri_f^{II}}\approx 0.2$ with a variability of $\pm 0.05$ within our experimental parameter range, which is marginally different from the generally accepted value of 0.17 used in oceanic models. The marginal difference could indicate that the mixing efficiency has a strong dependence on the mechanism by which turbulence is generated and the turbulence regime itself. Following this, $\overline{Ri_f^{II}}$ was used to parameterize the momentum eddy diffusivity, $\kappa_{m}$ and scalar eddy diffusivity, $\kappa_{\rho}$. It was observed that the eddy diffusivity of momentum was in good agreement (within our experimental parameter range) with previous literature but the eddy diffusivity of scalar showed an agreement up until $Re_{b}<$ 10 (the diffusive regime). The departure in the scalar diffusivity values in the intermediate regime, $Re_{b}>$ 10, suggested the importance of quantifying the small-scale mixing dynamics of an energetic gravity current. We believe that these results will provide good insights on the small-scale local dynamics and energetics of a lock-exchange gravity current with turbulence regime covered here being relevant to oceanic flows as well.

\begin{acknowledgments}
All the experiments reported in this study were conducted at University of Notre Dame, USA in the Department of Civil and Environmental Engineering and Earth Sciences when the corresponding author (SB) was at University of Notre Dame as a Visiting Professor. SB is grateful to Prof. H.J.S Fernando (Wayne $\&$ Diana Murdy Endowed Professor, University of Notre Dame) for giving access to the experimental facility and providing useful inputs for this work.
\end{acknowledgments}

\bibliography{apssamp}

\end{document}